\documentclass{elsarticle}

\usepackage{graphicx}
\usepackage{comment}
\usepackage{url}
\usepackage{adjustbox}

\usepackage{xcolor}

\newif\ifdraft
\drafttrue

\journal{Systems and Software}

\begin{document}

\begin{frontmatter}


\title{An analysis of open source software licensing questions in Stack Exchange sites
}
\author[1]{Maria Papoutsoglou\corref{cor1}}
\ead{mpapouts@csd.auth.gr}
\author[2]{Georgia M. Kapitsaki\corref{cor1}}
\ead{gkapi@cs.ucy.ac.cy}
\author[3]{Daniel German}
\ead{dmg@uvic.ca}
\author[1]{Lefteris Angelis}
\ead{lef@csd.auth.gr}

\cortext[cor1]{Corresponding author}

\address[1]{Aristotle University of Thessaloniki, Thessaloniki, 541 24, Greece}
\address[2]{University of Cyprus, 1, University Avenue, Aglantzia, 2109, Cyprus}
\address[3]{University of Victoria, Victoria, BC V8P 5C2, Canada}

\begin{abstract}
Free and open source software is widely used in the creation of software systems, whereas many organisations choose to provide their systems as open source. Open source software carries licenses that determine the conditions under which the original software can be used. Appropriate use of licenses requires relevant expertise by the practitioners, and has an important legal angle. Educators and employers need to ensure that developers have the necessary training to understand licensing risks and how they can be addressed. At the same time, it is important to understand which issues practitioners face when they are using a specific open source license, when they are developing new open source software products or when they are reusing  open source software. In this work, we examine questions posed about open source software licensing using data from the following Stack Exchange sites: Stack Overflow, Software Engineering, Open Source and Law. We analyse the indication of specific licenses and topics in the questions, investigate the attention the posts receive and trends over time, whether appropriate answers are provided and which type of questions are asked. Our results indicate that practitioners need, among other, clarifications about licensing specific software when other licenses are used, and for understanding license content. The results of the study can be useful for educators and employers, organisations that are authoring open source software licenses and developers for understanding the issues faced when using licenses, whereas they are relevant to other software engineering research areas, such as software reusability.
\end{abstract}

\begin{keyword}
open source software \sep Stack Exchange \sep software licenses \sep topic modeling
\end{keyword}

\end{frontmatter}

\section{Introduction}

Open Source Software (OSS), particularly in the form of libraries and frameworks, has become a fundamental part of software development and is widely used in research; almost any software today relies on some OSS \cite{synopsys2020,sonatype2020}. Reusing OSS creates a variety of legal concerns for software developers and their employers. Firstly, a component's license might impose restrictions on whether and how the software component is incorporated into a system; for example, unless adequately interconnected with the rest of the system, using a component licensed under the GNU General Public License (GPL) might severely restrict the ability to commercialise a product built upon it. Secondly, if the organisation incorporates OSS in the software that it publishes (whether for profit or not) it needs to know what requirements it must satisfy (these requirements are imposed by the software licenses of the components that it incorporates). Moreover, in software engineering research, OSS licensing is an important parameter in areas, such as project management, requirements and reusability~\cite{haefliger2008code}. Thus, adopting an OSS component requires both an informed process to evaluate the legal ramifications of using the component and a process to ensure that the component is used according to its license (license compliance).  

Software developers, as part of their day-to-day activities are responsible, at least in part, for both of these tasks (adoption and license compliance). Given the proliferation of OSS reuse, they are required to have a basic understanding of OSS licensing, at the very least to know when to ask for legal advice. As a consequence, educational institutions and employers need to make sure that software developers have the necessary training to be aware of the risks associated with licensing and how to address or mitigate them.

Questions-and-answers (Q\&A) sites, such as Stack Overflow\footnote{\url{https://stackoverflow.com/}} (SO), have become a major source of information for software developers. These sites provide a valuable window into the challenges that software developers face. On the one hand, they provide a list of questions that developers have; because this list is ranked based on popularity (e.g. Stack Overflow up votes) these questions can be used to understand what are the most common questions or problems that developers have. On the other hand, the answers provided to these questions can be used to ascertain how much agreement or disagreement exists among people who answer these questions. Previous works have analysed the content of discussions on Stack Exchange sites, in order to examine the questions posts in a specific domain~\cite{abdellatif2020challenges,ahmed2018concurrency}. This is also the case for questions regarding OSS licensing. This information can highlight specific topics that must be addressed in training and in organisation's policies that describe how OSS should be adopted and managed.

Previous empirical studies have examined license usage in the framework of specific projects, communities or using the history of projects on GitHub~\cite{di2010exploratory,moraes2021one,vendome2017license}. However, the directions training activities need to be directed towards concerning licensing have not been examined. In this paper, we have studied the questions and answers related to OSS licensing and reuse in four Q\&A sites: Stack Overflow, Law\footnote{\url{https://law.stackexchange.com/}}, Open Source\footnote{\url{https://opensource.stackexchange.com/}} (OS) and Software Engineering\footnote{\url{https://softwareengineering.stackexchange.com/}} (SEng). From these sites we collected 6,697 questions regarding OSS licensing and their 11,691 answers. We focused our study on the following Research Questions (RQs):
\begin{itemize}
\item \emph{RQ1: Which OSS licenses do users mention in their questions and how do they evolve over time?}: we have identified the top licenses appearing in question posts and have found that users' questions contain a larger number of licenses in recent years, whereas GPL, MIT and LGPL (GNU Lesser General Public License) are the licenses present most frequently. 
\item \emph{RQ2: Which are the most frequent topics appearing in OSS licensing questions?}: we have identified the most common topics that appear in questions and have seen that users ask most questions about specific licenses and the distinction between commercial and OSS, whereas other specific topics include license conditions, software modifications, linking and repositories.
\item \emph{RQ3: Which questions on OSS licensing get more attention and which seem more difficult to answer?}: we have investigated the popularity and difficulty of questions by studying questions in relation to the answers they receive and have found that questions that refer to LGPL-3.0, EPL (Eclipse Public License) and BSD-3-Clause take longer to get an accepted answer, whereas questions on linking and selling software do not have an accepted answer in many cases.
\item \emph{RQ4: Which type of questions do OSS users ask about licensing?} (e.g. how, what): we have seen that most user questions do not correspond to common question types.
\end{itemize}

These results point to directions for improvement in education and organisational practices. They can be a guide for educators in training engineers better on open source software and licensing issues. Lawyers and organisations that might create or maintain license texts and documentation about licenses can understand on which parts of the documentation they should put emphasis. Developers creating open source software could be better directed in understanding licensing issues faced by other developers.

An initial analysis on licensing questions and answers texts using 3 Stack Exchange sites has been performed in our previous work~\cite{kapitsaki2020developers}. In this work, we extend our dataset by considering Stack Overflow as an additional source and we investigate new research questions but with stricter criteria for inclusion in the dataset. We are introducing new areas of analysis, capturing whether practitioners receive an appropriate response in their questions, we analyse the type of questions asked and examine trends in topics over time.

\section{Related work}
\label{relatedwork}

\subsection{Previous works on Licensing}

Many previous works have focused on licensing issues in open source software. License identification tools, such as FOSSology~\cite{gobeille2008fossology}, Ninka~\cite{german2010sentence}, the Open Source License Checker (OSLC)~\cite{xu2010design} and the Automated Software License Analysis (ASLA)~\cite{tuunanen2009automated}, rely mainly on regular expressions and patterns stored in a database, in order to extract information on license(s) using as input the software source code. Relevant information on additional license identification techniques can be found in a previous survey~\cite{kapitsaki2015insight}. 

Although not adequate emphasis was given by software engineers and social coding platforms on license selection in the past, there is currently a tendency of providing mechanisms for better informed users. For instance, GitHub gives users the possibility to select among a number of licenses via \emph{choosealicense}\footnote{\url{https://choosealicense.com/}}, and \emph{tldrLegal}\footnote{\url{https://tldrlegal.com/}} provides information about each license, so that users are aware of what they can, cannot and must do with every license, whereas \emph{findOSSLicense} is a license recommender~\cite{kapitsaki2019modeling}. License modeling has also been examined in the past. Graph approaches, where licenses are denoted as nodes and the compatibility between licenses is expressed with edges, can be encountered, such as the license slide of Wheeler~\cite{wheeler2007free} and the license compatibility graph~\cite{foukarakis2012choosing}. 

A survey performed with contributors of 104 popular GitHub systems examined usage license declarations as one of the factors that are relevant to the success or failure of an open source project on GitHub~\cite{coelho2017modern}. Areas of improvement in developers' knowledge about licenses were identified in a study that examined whether developers understand open source licensing~\cite{almeida2017software}. 42 different cases of using code under different open source licenses, i.e. GPL-3.0, LGPL-3.0 and MPL-2.0 (Mozilla Public License 2.0), were given and developers were asked how they would behave. The main conclusions were that developers were able to interpret a variety of simple and complex development scenarios but they do not have a consistent understanding of what technical details matter and how licenses interact. 

The usage of licenses in the JavaScript community, and specifically how multi-licensing is used, is studied in~\cite{moraes2021one}. Multi-licensing in this work is viewed as the case where one project has components that are licensed under different licenses. The authors found that the majority of JavaScript projects use more than one license, whereas a small part of those projects uses also non-software licenses. The usage of licenses that are not approved by the Open Source Initiative (OSI) is examined in~\cite{meloca2018understanding}. Different versions of open source software projects were studied, whereas the developers of some of these projects were also contacted. Many cases, where a software license would be changed from a non-approved to an approved license were found. 

The licensing evolution of six open source systems was studied in~\cite{di2010exploratory}. The reasons why developers change the license of their projects using the commit history of Java projects and a developers survey were investigated in~\cite{vendome2015and}. 1,200 documents from issue trackers and legal mailing lists were used in order to devise a catalog of licensing bugs and how they can categorised (e.g. community guidelines, what is derivative work, what is copyrightable, incorrect licensing)~\cite{vendome2018distribute}. A work on license usage on GitHub based on data available in 16,221 Java projects analysed change patterns in licensing~\cite{vendome2017license}. Its main results were that: developers adopt new license versions quickly, license changes move mainly toward permissive licenses, developers discuss licensing in the issue tracker of their project, and there is no standardization or consistency on how a license is attributed to a software system.
 
\subsection{Studies on Q\&A platforms}

Regarding the usage of source code for the case of Stack Overflow, a previous work examined the issue in missing attributions when copying code snippets that may lead to licensing issues when pasting code from Stack Overflow in GitHub projects~\cite{baltes2019usage}. Stack Exchange and its Q\&A sites have been used however, for different kinds of analysis in different domains. Works in the security~\cite{yang2016security,lopez2019anatomy} and testing domain~\cite{kochhar2016mining} can be encountered, with all works using data from Stack Overflow. Stack Overflow was also used in order to examine the topics chatbot developers are interested in and the challenges they face~\cite{abdellatif2020challenges}. The authors examined the topic found in posts' titles, the types of questions divided into how, why and what, and which topics are most difficult to answer, whereas they also performed an analysis on the evolution of topics over time. Similar analysis has been performed for Docker
~\cite{haque2020challenges}, code smells and anti-patterns~\cite{tahir2018can} and concurrency development~\cite{ahmed2018concurrency}.
\\
\\
\textbf{Comparison to previous works:} In relevance to other examinations of Q\&A posts, we focus on an aspect that has not been investigated, i.e. licensing in open source software, whereas we consider additional relevant Stack Exchange sites, apart from Stack Overflow. This way, we are providing a view from different perspectives (e.g. software development, legal). In relevance to prior works on licensing, no similar study exists, to the best of our knowledge. Since we are not focusing on specific open source communities, we are offering a more generic view on aspects practitioners face when dealing with open source software and its licensing.

\section{Research questions}
\label{rqs}

In more detail, the introduced Research Questions are the following:
\begin{itemize}
    \item \textbf{RQ1}. \emph{Which OSS licenses do users mention in their questions and how do they evolve over time?}\\
    This research question aims in identifying licenses and their versions in the question texts (e.g. GPL-2.0). We study which specific licenses appear in user questions in each site, and whether there are any changes over time, whereas we are also examining differences between the sites. This assists in understanding the licenses the practitioners are mentioning more, which may mean that they find these licenses more often in third party software they are using or that they have more difficulty in understanding them. The evolution over time is useful in order to observe any changes in the number and type of licenses that appear in questions (e.g. there is generally a recently observed tendency toward permissive licenses~\cite{vendome2017license}).
    \item \textbf{RQ2}. \emph{Which are the most frequent topics appearing in OSS licensing questions?} \\
    Licensing questions concern different subjects and may include clarifications about specific licenses, or clarifications on general licensing concepts, such as linking. For this research question, we examine the topics that appear more frequently using topic modeling techniques, and specifically Latent Dirichlet Allocation (LDA). As part of the discussion on topics, we investigate changes in topics distribution over time.
    \item \textbf{RQ3}. \emph{Which questions on OSS licensing get more attention and which seem more difficult to answer?}\\
    In this research question, we use a number of metrics that help us understand better the attention questions receive and the characteristics of responses. We examine the following aspects: average views, average favourites, average scores, posts without an accepted answer, average number of answers and median time to get an accepted answer. This analysis is performed using the license names and versions, and also the licensing topics. 
    \item \textbf{RQ4}. \emph{Which type of questions do OSS users ask about licensing?}\\
    We examine the type of question: what, how, why, which, other, relying on topic modeling results. This analysis can help us draw more specific conclusions about the kind of questions practitioners usually have. 
\end{itemize}

\section{Study methodology}

\begin{figure*}[!t]
\centering
\includegraphics[scale=0.44]{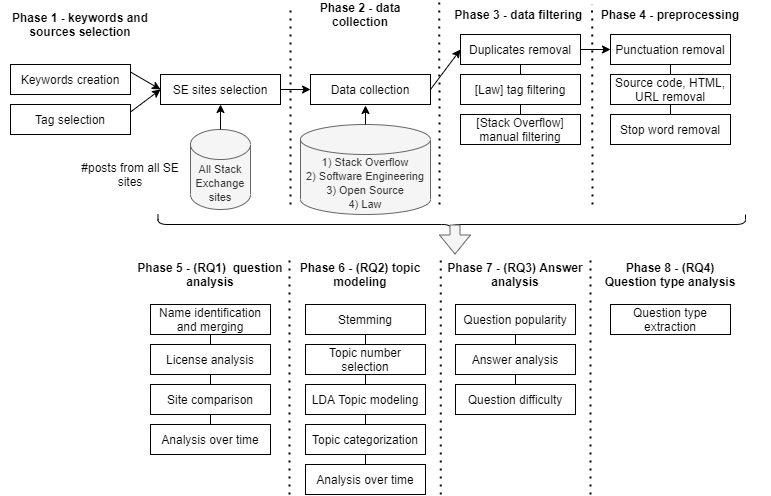}
\caption{Analysis phases.}
\label{fig:phases}
\end{figure*}

\subsection{Keywords, sources selection and data collection (Phases 1-2)}

The phases that we have followed for our analysis are depicted in \figurename~\ref{fig:phases}. We have relied on both keywords and tags in order to collect data about licensing from different Stack Exchange sites. We have used the following process in order to choose which sites to include in the study, as more relevant to licensing (Phase 1):
\begin{enumerate}
    \item We have created keywords related to licensing, adopting initially keywords used in the study of~\cite{vendome2017license}, excluding however, the following keywords that are very generic and point to discussions irrelevant to licensing in many Stack Exchange sites, as we manually verified by inspecting the text of questions returned in Stack Exchange online search\footnote{\url{https://stackexchange.com/search}}: \emph{merchantability}, \emph{written permission}, \emph{prior permission}, \emph{see the copyright.txt}, \emph{liability}, \emph{legal}, \emph{special exception}, \emph{to permit this exception}. We randomly sampled data from 20 questions for each keyword and when all of them were irrelevant to licensing we ignored the keyword (this process was performed by one of the authors).
    \item We created additional keywords by adding names of licenses appearing differently, as we expect to find differences in how users express a license name (e.g. CPOL 1.02, CPOL-1.02, CPOL v1.02, CPOL-v1.02 all refer to the same license). 
\end{enumerate}

The final list of keywords used in all sources as we describe next (except Stack Overflow site after the final Stack Exchange sites selection) is found in Table~\ref{tab:keywords}. In addition to those keywords, we used full license names that are not already part of the keywords as appearing in the SPDX (SPDX (Software Package Data Exchange) license list, e.g. Boost Software License 1.0~\cite{stewart2010software}. We do not list the SPDX license list here as it is available online\footnote{\url{https://spdx.org/licenses/}}.

\begin{table*}
  \caption{Keywords employed for collecting license-relevant questions texts}
  \label{tab:keywords}
  \begin{adjustbox}{width=0.98\textwidth}
  \begin{tabular}{lll}
    \hline
MIT license, 
General Public License, 
General Public License v2, 
General Public License-2, 
General \\Public License-2.0, 
General Public License v2+, 
General Public License-2+, 
General Public \\License-2.0+, 
General Public License 2, 
General Public License 2.0, 
General Public License 2+, \\
General Public License 2.0+, 
General Public License v3,  
General Public License-3, 
General \\Public License-3.0,
General Public License v3+, 
General Public License-3+, 
General Public\\ License-3.0+, 
General Public License 3, 
General Public License 3.0,
General Public License 3+, \\
General Public License 3.0+, 
Apache License 2.0, 
Apache License-2.0, 
Apache License 2, 
Apache \\License-2, 
Apache License v2, 
Apache License-v2, 
Apache License v2.0,
Apache License-v2.0, 
Apache \\License 1.1, 
Apache License-1.1, 
Apache License v1.1, 
Apache License-v1.1, 
BSD License 2.0, 
BSD \\License-2.0, 
BSD License 2, 
BSD License-2, 
Lesser General Public License 2.1, 
Lesser General Public\\ License v2.1, 
Lesser General Public License-2.1, 
Lesser General Public License v2.1+,
Lesser General \\Public License-2.1+, 
Lesser General Public License 3.0, 
Lesser General Public License v3.0, 
Lesser \\General Public License-3.0, 
Lesser General Public License v3.0+, 
Lesser General Public License-3.0+, \\
Eclipse Public License, 
Microsoft Public License, 
Simplified BSD License, 
Code Project Open License \\1.02, 
Code Project Open License, 
Mozilla Public License 1, 
Mozilla Public License 1.0, 
Mozilla Public \\License v1.0, 
Mozilla Public License-1, 
Mozilla Public License-1.0, 
Mozilla Public License v1, 
Mozilla \\Public License 2, 
Mozilla Public License 2.0, 
Mozilla Public License v2.0, 
Mozilla Public License-2, \\
Mozilla Public License-2.0, 
Mozilla Public License v2, 
Affero General Public License v3, 
Affero General \\Public License 3, 
Affero General Public License 3.0, 
Affero General Public License-v3, 
Affero General \\Public License-3, 
Affero General Public License-3.0, 
Common Development and Distribution License, \\
DO WHAT THE F*CK YOU WANT TO PUBLIC LICENSE, 
Microsoft Reciprocal License, 
zlib/libpng \\License, 
ISC license, 
artistic license, 
artistic 1.0, 
artistic-1.0, 
artistic 2.0,
artistic-2.0, 
artistic v1.0, \\
artistic-v1.0, 
artistic v2.0, 
artistic-v2.0, 
MS-PL, 
MS-RL, 
CPOL 1.02, 
CPOL-1.02, 
CPOL v1.02,
license,\\
CPOL-v1.02, 
AGPL 3, 
AGPL v3, 
AGPL-3, 
AGPL-v3, 
zlib/libpng,
licensing, 
licensed, 
licenses, 
licensing,\\
disclaims copyright, apl-1.1, apl 1.1, epl-1, epl 1, epl v1, mpl v1.0, mpl-1.0, mpl-1, mpl 1, mpl-2.0, mpl 2.0, \\
mpl v2.0, mpl v2, mpl-2, mpl 2, bsd 4-clause, bsd-4, epl v1.0, cpl-1, cpl-1.0, cpl 1.0, cpl 1, cddl v1.0, \\
cddl-1.0, cddl 1.0, mpl 1.0, apache-1.1, apache 1.1, epl-1.0, epl 1.0, lgpl, lgpl-2, lgpl 2, lgpl v2, lgplv2, \\
lgpl-3, lgpl 3, lgpl v3, lgplv3, lgpl v2.1, lgplv2.1, lgpl-2.1, lgpl 2.1, bsd-2, bsd\_3-clause, bsd-3, apache v2, \\
apache v2.0, apache-2.0, apache 2.0, apache-2, apache 2, gpl, gpl-3, gpl-2, gpl 3, gpl 2, gpl3, gpl2, gpl v2, \\
gpl v3, gplv2, gplv3, public domain\\
\hline
\end{tabular}
\end{adjustbox}
\end{table*}

We used the keyword list created by the above process to collect how many relevant questions exist in each Stack Exchange site, i.e. questions that refer to open source software and its licensing, using Stack Exchange search. We searched specifically for each keyword in the question title and body. Based on the response, we selected the following sources for the final data collection that appeared in the top 6 positions in the posts returned: 1) Stack Overflow (in position 1 with the highest number of questions containing our keywords: 12,980 questions), 2) Open Source (in position 2: 2,347 questions), 3) Software Engineering (in position 4: 1,424 questions) and 4) Law (in position 6: 892 questions). The Server Fault and Super User websites also appeared in the list (in positions 3 and 5, and with number of questions 2,148 and 954 respectively), but they were both discarded after manual inspection, since indications of license names in the posts texts refer to software products and not to OSS licenses. 20 question texts from each site were read by two of the authors before taking this decision and they were all not referring to licensing aspects. Some examples of questions that we encountered and were not relevant are: Super User \#784244: ``\emph{Don't have access to download files on Apache 2}" referring to Apache server, and Server Fault \#341914: ``\emph{FreeBSD out of box Screen/TMUX functionality?}," mentioning BSD-3-Clause in the question body to make a clarification on licensing but the question author asks about differences between tools and not about licensing. 

Stack Overflow is a special case, so we followed a different approach to select data. Stack Overflow appeared first in the list, but licensing questions are outside its scope, as indicated in the directions provided for the users in the tag descriptions that refer to open source software licensing. For instance, the \emph{gpl} tag description indicates that GPL license questions are out of topic: ``\emph{DO NOT USE! LICENSING / LEGAL ADVICE IS OFF TOPIC -- GPL questions should be asked on opensource.stackexchange.com - The GNU General Public License is a copyleft free software license and is the most popular open source license}." This is also the case for the \emph{lgpl} and \emph{licensing} tags, whereas other tags with license names refer to software products and not to the respective license (e.g. the \emph{zlib} tag is described as: ``\emph{zlib is a library used for data compression})." According to Stack Exchange policies questions that are not appropriate for a specific Stack Exchange site are migrated\footnote{\url{https://meta.stackexchange.com/questions/10249/what-is-migration-and-how-does-it-work}} to the suitable site. 

This issue becomes more apparent, when questions that contain the generic term \emph{license} are inspected, as the questions refer to applying the license of a specific software, such as activating a license key. Some representative question titles for the `\emph{license}' term are the following:
\begin{itemize}
    \item ``\emph{Visual studio community: License expired and License error (0x00000010): Cached online license container is corrupt.?}" (SO \#32413780)
    \item ``\emph{Oracle Mobile Application framework license}" (SO \#42600786)
    \item ``\emph{InnoSetup license files}" (SO \#12594634)
\end{itemize}

Nevertheless, a manual inspection that we performed on recent posts with the \emph{gpl} tag reveals that a number of Stack Overflow questions are relevant to licensing. For this reason, we decided to rely only on some licensing-relevant tags to collect data from Stack Overflow, as shown in Table~\ref{tab:collection-tags}. We also used licensing relevant tags from the other three Stack Exchange sites for additional data collection of posts that were not in the keyword-based collected data (Table~\ref{tab:collection-tags}), in order to ensure that all data referring to OSS licensing are collected. In order to select the tags, we manually examined all tags available in each site. 

Relying nevertheless, the data collection only on tags can prove less effective, as previous works have noted~\cite{barua2014developers,tahir2020large}. We observed some cases, where users are not adding the appropriate tags in their posts, even though they are relevant to licensing (e.g. OS \#468 with title: ``\emph{Will open source licenses enable me to choose the businesses model I want?}" and part of the body ``\emph{[...]By choosing an open source license, will this enable me to choose? Or will it be like copyleft businesses that are not free to choose?[...]}" has only the \emph{business} tag). We thus, decided to use the above hybrid approach of using both keywords and tags, in order to ensure that all relevant posts are included in our data. The process that we described above is depicted in Phases 1 and 2 in \figurename~\ref{fig:phases}.

\begin{table}
\centering
  \caption{Licensing relevant tags used per site and frequency in questions in the site (\#Q).}
  \label{tab:collection-tags}
\begin{adjustbox}{width=0.9\textwidth}
  \begin{tabular}{lclclclc}
    \hline
    \textbf{SO} & \textbf{\#Q} & \textbf{SEng} & \textbf{\#Q} & \textbf{OS} & \textbf{\#Q} & \textbf{Law} & \textbf{\#Q}\\
    \hline
licensing &	2,005&	licensing&	1,335&	licensing&	885	&licensing&	567\\
gpl	&338&	gpl	&497&	gpl&	672	&open-source-software&	211\\
lgpl	&118&	mit-license&	157&	mit&	309&	gpl	&78\\
&&lgpl&	122&	license-compatibility&	293\\
&&bsd-license &	101&	gpl-3&	281\\
&&apache-license&	97&	license-recommendation&	262\\
&&intellectual-property&25&apache-2.0&215\\
&&&&lgpl&	199\\
&&&&commercial&	180\\
&&&&derivative-works&	169\\
&&&&gpl-2&	152\\
&&&&license-notice&	151\\
&&&&agpl-3.0&	146\\
&&&&relicensing	&132\\
&&&&bsd&	129\\
&&&&law	&122\\
&&&&distribution&	121\\
&&&&copyleft&	111\\
&&&&attribution	&101\\
&&&&multi-licensing&	84\\
&&&&license	&76\\
&&&&mpl	&49\\
&&&&redistribution&	45\\
&&&&license-file&	44\\
&&&&public-domain&	40\\
&&&&epl	&17\\
&&&&permissive	&13\\
&&&&cddl&	12\\
&&&&ms-pl&	9\\
&&&&dual-license&	8\\
&&&&eupl&	8\\
&&&&zlib&	7\\
&&&&license-comparison&	6\\
&&&&bsl&	5\\
&&&&lppl&	5\\
&&&&unilicense&	5\\
&&&&agpl&	4\\
&&&&license-creation&	4\\
&&&&osl-3.0&	4\\
&&&&wtfpl&	4\\
&&&&0bsd&	3\\
&&&&cecill&	3\\
&&&&zlib/libpng	&3\\
&&&&apache-1.1&	2\\
&&&&afl	&1\\
&&&&apsl&	1\\
&&&&artistic-license-2.0&	1\\
&&&&cpal&	1\\
&&&&cpol&	1\\
&&&&isc	&1\\
&&&&open-game-license&	1\\
&&&&sgi-b&	1\\
\hline
\end{tabular}
\end{adjustbox}
\end{table}

\subsection{Dataset filtering (Phase 3)}

Since some posts may have been collected more than once (when the question contains both tags and keywords of the collection process), we removed duplicates using the question id for every site and the body text (Phase 3 in \figurename~\ref{fig:phases}). This ensured also the removal of duplicate questions across the Stack Exchange sites, i.e. questions that were posted to more than one sites at the same time (we encountered 28 such cases). 

A separate filtering was performed for the case of Law Stack Exchange site. 
With manual inspection, we saw that many questions are referring to copyright law outside software (e.g. videos). We created a set of relevant tags and filtered out the respective questions (e.g. \emph{photography, social-media, youtube, music, online-piracy, tax-law, corporate-law, labor-law}). We have kept copyright law questions that refer to patents or trademarks, as they are relevant to licensing, e.g. in the text of GPL-3.0 it is stated that: ``\emph{Each contributor grants you a non-exclusive, worldwide, royalty-free patent license under the contributor's essential patent claims.}" Questions that were referring to specific countries, or states (of United States) were also filtered out. Although the legislation on copyright law may differ between countries, it usually does not include software. For instance, the specific legal regime of open source software has not been explicitly considered in the course of reforms that were performed in the European legislation that concerns copyright law~\cite{synodinou2016cypriot}. 

Since the \emph{license} keyword is very generic and may be referring to areas outside OSS (e.g. license plates, license to practice a profession, license for buying a software), we decided to perform a manual filtering on the questions collected via the use of this keyword and that were not already included in the collected data via the other keywords and tag collection. We used the manual filtering as an additional process that would assist in improving the dataset quality, as a quality assurance step. This process was carried out by one of the authors and was verified by a second author. This was performed on the Stack Exchange sites apart from Stack Overflow: in the Software Engineering site, out of the 703 questions that did not contain other relevant keywords or tags, only 37 were kept (removed posts were referring mainly to license keys and using specific open source projects without reference to licensing); in the Open Source site, out of the 220 questions, 165 were kept (posts that were removed were referring, for instance, to open hardware); in Law all 1,268 posts were filtered out, as a manual inspection of 200 posts showed that none were relevant to software (the posts were referring to driver's license, license plates, image or product licenses).

For the case of Stack Overflow that contains as aforementioned a large number of irrelevant posts, we performed a manual filtering on the 2,193 question posts returned in total via the tag-based collection by inspecting the question title. Finally, 1,315 questions were included in the analysis dataset. Most of the questions removed were referring to license/product keys/numbers (e.g. how to create keys for software and how to use relevant libraries), paid licenses for specific software or the iOS developer license and were carrying the \emph{licensing} tag. The task was performed by one of the authors and the irrelevant posts identified were reviewed by a second author.

Our dataset was created in the second half of October 2020. It has 6,697 questions and 11,596 answers from the four Stack Exchange sites and the relevant answers after the respective filtering was applied (Table~\ref{tab:dataset-size}) and is available on Zenodo\footnote{\url{https://zenodo.org/record/4428267#.X_hW0OgzaUk}}. The questions from some sites have a smaller size than the dataset used in the existing work with the initial analysis~\cite{kapitsaki2020developers}, as in that work other general keywords were also used for data collection purposes and manual filtering was not performed.

\begin{table}
\centering
  \caption{Dataset size.}
  \label{tab:dataset-size}
\begin{adjustbox}{width=0.5\textwidth}
  \begin{tabular}{lcc}
    \hline
    \textbf{Stack Exchange site} & \textbf{Questions} & \textbf{Answers}\\
    \hline
    Stack Overflow & 1,295 & 2,862\\ 
    Software Engineering & 1,687 & 3,537\\ 
    Open Source & 3,030 & 4,353\\ 
    Law & 685 & 844\\ 
    \hline
    \textbf{Total} & 6,697 & 11,596\\ 
    \hline
\end{tabular}
\end{adjustbox}
\end{table}

\subsection{Preprocessing (Phase 4)}

A number of preprocessing steps were required, in order to prepare the data for the analysis (Phase 4 in \figurename~\ref{fig:phases}). The following generic preprocessing was performed to all text in the order indicated below.\\
\textbf{Removal of punctuation marks.} All punctuation marks were removed and all text was converted to lower case for easier manipulation and text comparison. \\
\textbf{Source code, HTML tags and URL removal.} A large number of questions and answers contains source code (indicated by the $<$code$>$ $<$/code$>$ tag). We removed source code, as code fragments are not needed in our analysis and may add noise. HTML tags (e.g., $<$li$>$, $<$p$>$ or $<$a$>$) and URLs were also removed, as they are not useful to our analysis and may add irrelevant data especially for the topic modelling phase.\\
\textbf{Stop word removal.} We removed stop words using a list of common English stop word list provided by the \emph{quanteda} (Quantitative Analysis of Textual Data) R package~\cite{benoit2018quanteda}. We also removed a number of dedicated stop words that we created. The list consists of the following words that are common in Q\&A websites, but do not add value in the analysis: \emph{data, question, answer, found, exception thrown, hope helps, throw exception, answer question, works fine, stack trace, code exception, error message, catch exception, exception occurs, solve problem, compiler compliance level, msdn.microsoft.com en-us library, version culture neutral, culture neutral publickeytoken}. The list was created by one of the authors and was reviewed by the other authors. 

Different preprocessing was performed in order to extract information on specific licenses and their versions from the question text (RQ1). \\
\textbf{License name and version identification.} To identify specific open source licenses, when looking in the questions text, we used a list that included the license names and versions used for the dataset collection including license names and their abbreviations as described in the SPDX. \\
\textbf{License name and version merging.} The same license may appear in different format in the Stack Exchange posts, e.g. a space or `-' may appear between the license name and its version, or the whole license name may be used instead of its abbreviation. We merged cases where the same license appears many times in the text, and we also merged cases where different terminology is used for the same license. For instance, we observed that the MIT license may appear in the text also as ``mit expat", whereas Microsoft Public License may appear also as MS-PL. We performed this merging, in order to be able to consider each license one time in each question text.\\
\textbf{License name with no version indication.} We kept cases of licenses with no version indication separately, as we observed many such cases. However, when a license with a version was also indicated in the same post, we kept only the license with version. We consider such license cases as a license family that is characterized by some main common features. We added this provision for the following license families: GPL, LGPL, AGPL, MPL, EPL, Apache, BSD.

In order to prepare the text for topic modelling linked with RQs 2 to 4, the following step was performed in addition to the generic preprocessing steps: \\
\textbf{Stemming.} Stemming leveraged from \emph{quanteda} R package was used to reduce the words to their root representations, for instance, \emph{licenses}, and \emph{licensing} both get stemmed to the word \emph{licens}.

The \emph{quanteda} R package was also used for the text analysis of the preprocessing phase. R was the programming language used also for the data  collection (Phase 2 in \figurename~\ref{fig:phases}), via the Stack Exchange API, and the analysis performed for the research questions of the study.

\section{Analysis and Results}

We have employed different techniques for the analysis required to answer our research questions. The steps followed for the results are depicted in Phases 5 to 8 in \figurename~\ref{fig:phases}. In all analysis phases, we used the question text, whereas we are also using the answer text in specific cases indicated.

\subsection{RQ1. Which OSS licenses do users mention in their questions and how do they evolve over time?}

In this research question we focused on examining which licenses are commonly found in users' posts, which differences can be observed between the sites, and how the appearance of license names in questions changes over time. The body of the question text was employed.

Table~\ref{tab:licenses-qa-sites-freqs} presents the license indication frequencies in questions per site. We are showing the 39 top appearing licenses in the Open Source site and their frequencies in the other 3 sites, as the Open Source site is the most appropriate site to pose questions on OSS licensing. GPL without version is the top appearing license family (22.9\% of questions) and its most appearances can be found in the Open Source site, whereas LGPL without version appears also among the top licenses in questions (4th top appearance or 7.8\% of questions). GPL licenses, with or without version, are indicated in 39\% of the questions. GPL3.0 appears in 8.9\% of the questions since 2008 (3rd top appearance overall), whereas GPL-2.0 appears in 6.7\% (5th top appearance overall). The second top license in questions is MIT appearing in 15.6\% of questions with a highest presence in the Open Source site. Apache-2.0 is the 6th top license indicated in 6.2\% of questions.

\begin{table*}
  \caption{Frequencies of licenses in questions.}
  \label{tab:licenses-qa-sites-freqs}
  \centering
  \begin{adjustbox}{width=0.9\textwidth}
  \begin{tabular}{ccccc}
    \hline
    \textbf{License}& \textbf{Stack Overflow} & \textbf{Software Engineering} & \textbf{Open Source} & \textbf{Law}  \\
     \hline
GPL	&372	&460	&631	&70\\
MIT	&83	&302	&591	&71\\
GPL-3.0	&52	&146	&367	&28\\
GPL-2.0	&71	&95	&265	&19\\
Apache-2.0	&37	&94	&260	&22\\
LGPL	&167	&151	&195	&12\\
AGPL-3.0	&17	&58	&159	&7\\
public-domain	&14	&40	&81	&40\\
BSD-3-Clause	&0	&18	&66	&3\\
BSD	&24	&65	&64	&9\\
LGPL-2.0	&12	&17	&50	&3\\
LGPL-3.0	&7	&22	&46	&5\\
Apache	&18	&40	&45	&5\\
LGPL-2.1	&8	&15	&40	&1\\
MPL	&9	&22	&34	&0\\
MPL-2.0	&0	&5	&33 &0\\
BSD-2-Clause	&1	&14	&23	&2\\
EPL	&7	&13	&22	&2\\
WTFPL	&2	&9	&21	&6\\
zlib/libpng	&8	&6	&19	&2\\
Unlicense	&0	&5	&18	&3\\
GPL-3.0+	&0	&3	&16	&0\\
CDDL-1.0	&8	&6	&16	&1\\
GPL-2.0+	&1	&3	&12	&1\\
X11	&4	&2	&10	&1\\
MS-PL	&10	&15	&9	&2\\
EPL-1.0	&3	&1	&7	&0\\
CC.BY.4.0	&0	&0	&6	&0\\
X0BSD	&0	&0	&4	&1\\
CC.BY.SA.4.0	&0	&0	&4	&2\\
LGPL-2.1+	&0	&0	&4	&0\\
OFL-1.1	&0	&0	&3	&0\\
OpenSSL license	&0	&0	&3	&0\\
MPL-1.0	&4	&3	&3	&0\\
EPL-2.0	&0	&1	&3	&0\\
CPOL-1.02	&5	&3	&3	&1\\
Artistic-2.0	&3	&1	&3	&0\\
AFL-3.0	&0	&1	&3	&0\\
BSL-1.0	&0	&2	&3	&0\\
 \hline
\end{tabular}
\end{adjustbox}
\end{table*}

\begin{figure*}[!t]
\centering
\includegraphics[scale=0.52]{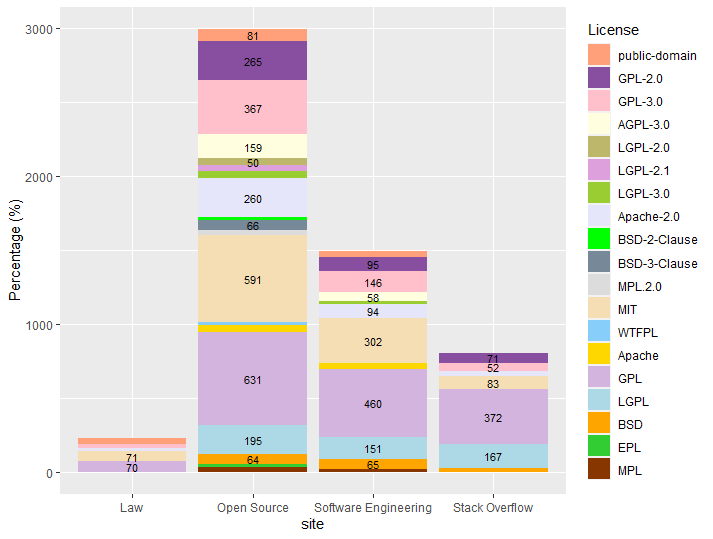}
\caption{Licenses asked about per Stack Exchange site.}
\label{fig:licenses-qa-sites}
\end{figure*}

In order to visualize the difference in the license presence in questions in the different sites, we present the top licenses appearing in user questions
in~\figurename~\ref{fig:licenses-qa-sites}. For readability purposes, we have omitted from the diagram licenses that appear less than 20 times in the respective site. The most diverse set of licenses is found in the Open Source site. In Law, the different licenses we find in the questions are limited. The most dominant licenses are 5 (MIT, GPL, public-domain, GPL-3.0, Apache-2.0) including the public-domain, that is plausible considering that it is more generic and may capture cases of non-software licensing. 

In order to examine the evolution of top licenses over the years, we show in Table~\ref{tab:dataset-size-peryear} the number of questions and answers in our dataset per year. Stack Overflow is the only site with questions with references to licenses in 2008. Questions in the Open Source site that has the largest proportion of questions in our dataset appear since 2015 and this explains the increase in the appearances since 2015. There is also an increase in the number of different licenses that appear in questions since 2015: the average number of different licences that appear per year in questions in the period 2008-2014 is 25.1, whereas in the period 2015-2020 it rises to 42.8.

\begin{figure*}[!t]
\centering
\includegraphics[scale=0.56]{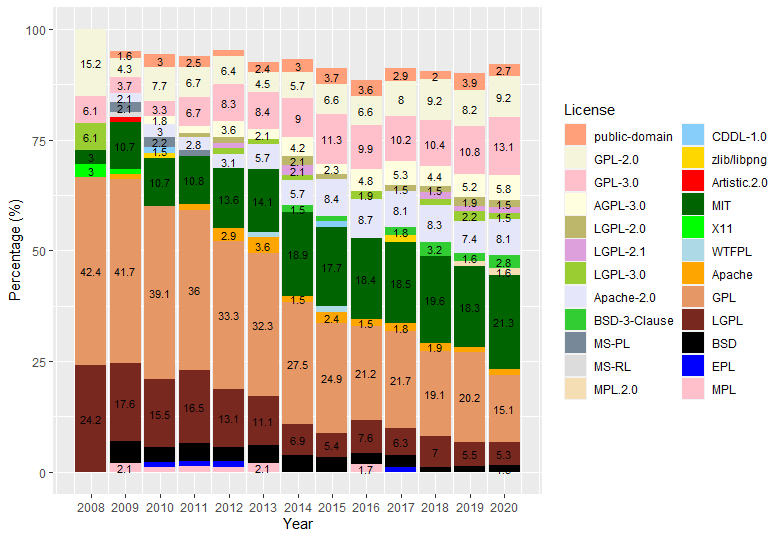}
\caption{Licenses appearing in questions over time.}
\label{fig:licenses-q-peryear}
\end{figure*}

The percentage of licenses that appear in questions over time are shown in~\figurename~\ref{fig:licenses-q-peryear}. The graph contains only licenses that appear in more than 1\% of the questions asked in total in all years. GPL without version is as aforementioned the case encountered more frequently in questions, but its presence has a decrease in the last years (its average appearance over the years is 28.8\% but only 15.1\% in 2020). LGPL presence also decreases over the years that might mean that users do not encounter the license family that often any more and do not ask therefore, about it. There is an increase in the appearance of the permissive MIT license since 2015 (MIT appears 63 times in 2014 and 116 times in 2015, whereas it has risen to 158 in 2020). Posts that contain reference to the public-domain have an appearance in all years.

\begin{table}
  \caption{Questions, answers and different licenses in all sites per year.}
  \label{tab:dataset-size-peryear}
  \centering
  \begin{adjustbox}{width=0.99\textwidth}
  \begin{tabular}{cccccccccccccc}
    \hline
     & 2008 & 2009 & 2010 & 2011 & 2012 & 2013 & 2014 & 2015 & 2016 & 2017 & 2018 & 2019 & 2020\\
     \hline
    \#questions & 46& 216  & 295 & 563  &499 & 385 & 319 & 835 & 836&677 & 616 & 679& 731 \\
    \hline
    \#answers to & 214&  685 &  751&  1,443 & 1,073&  714&  531&  1,440& 1,244& 901& 821 & 868&  911 \\
    questions &\\
    \hline
    \#different  & 8 &24 &26 &32 &28 &29 &29 &42 &42 &41 &43 &47 &42 \\
    licenses in &\\
    questions &\\
  \hline
\end{tabular}
\end{adjustbox}
\end{table}

\subsection{RQ2. Which are the most frequent topics appearing in OSS licensing questions?}

For the topics analysis we have employed the LDA topic modelling algorithm, as it is considered to provide better results than other approaches and it has been used in a variety of previous empirical studies in software engineering~\cite{blei2003latent,abdellatif2020challenges}. We performed trimming and removed specifically words that appear only 10 times or less in the text. 

\textbf{Number of topics.} The number of topics $K$ to be used in topic analysis is an important factor that needs to be determined. A large number of topics may provide very specialised results, whereas a low number may not be able to capture appropriately all the topics that appear in the questions. For this purpose, we employed the $C_v$ coherence metric to assist in the selection of $K$. 
LDA was run 3 times with coherence metric, in order to guarantee the accuracy of the results. We observed that the coherence metric does not differ a lot for $K$=12 and $K$=16 number of topics: the value was 0.59 and 0.58 respectively approximately. We manually verified that 16 topics is the ideal number, as all topics point to licensing aspects and, with minor exceptions, all topics refer to different aspects indicated with the presence of different terms that describe the topic. Based on this process, the final number of topics was set to $K=16$.

The topic analysis was performed on the questions, considering the text of the question body. We followed a different approach than the initial publication where both questions and answers text was analysed together, in order to be able to focus on the practitioner's questions and discuss the topics appearance in relation to the answers received~\cite{kapitsaki2020developers}. 

\textbf{Implementation of LDA.} For implementation purposes, relevant packages of the R programming language were used for the LDA implementation with the \emph{stm} as main package~\cite{roberts2019stm}. 

The 16 topics with the assigned names are presented in Table~\ref{tab:qa-topics}, along with the top 15 terms that appear in each topic and a category that was manually assigned, in order to group some topics together, as they were referring to close concepts of licensing. We used 2-gram and 3-gram for the term creation, as this way we can observe groups of words that form small phrases and this makes the topic terms more relevant to open source licensing and allows drawing more conclusions. The topic terms appear in their stemmed version. We have removed n-grams with words that repeat themselves, e.g. when ``permiss notic includ" and ``permiss notic" both appear as terms in the topic, we are keeping only one of them.

\noindent\fbox{%
    \parbox{\columnwidth}{%
The questions on licensing can be grouped in the following 7 main categories: \emph{Specific licenses},  \emph{Conditions}, \emph{Commercial vs. OSS}, \emph{Modifications}, \emph{Linking}, \emph{Repositories} and \emph{General OSS}. 
    }%
}

\begin{table}
  \caption{License topics, category and most frequent terms appearing in questions.}
  \label{tab:qa-topics}
  \begin{adjustbox}{width=0.99\textwidth}
  \begin{tabular}{cccl}
  \hline
    \textbf{Category} & \textbf{Topic \#} & \textbf{Topic name} & \textbf{Terms}\\
    \hline
Specific & 1& MIT \& &
mit licens, 
copyright notic, 
licens mit, 
subject condit, licens\\

licenses && notices & 
mit licens, 
substanti portion, 
permiss notic includ, copyright\\
&&& notic permiss, 
copi substanti portion, 
notic includ copi, includ copi\\
&&& substanti, 
notic permiss notic
notic includ, 
includ copi, 
code mit
\\

\hline
Conditions & 2& License &
copyright notic list, 
notic list condit, 
list condit disclaim, 
warranti
\\

&& conditions &
merchant fit, 
list condit, 
notic list, 
aris connect, 
softwar eal, \\
&&&
contract tort aris, 
merchant fit purpos, 
copyright holder liabl, 
\\
&&&holder liabl, 
liabl claim damag,
permit person softwar
\\

\hline
Specific & 3& GPL &
gnu gpl, 
gpl v3, 
dual licens, 
gpl version, 
driver licens, 
gnu gpl  \\
licenses&&&
licens,
licens gnu, 
chang licens, 
communiti edit, 
commerci softwar, \\
&&&gnu gpl v2, 
gnu gpl v3, 
softwar version, 
gpl v3 licens, 
licens fee
\\

\hline
Commercial& 4& Commercial &
close sourc, 
commerci licens, 
web applic, 
commerci applic, sourc \\
vs. OSS&&software &
 applic,
googl map, 
map api, 
close sourc applic, 
applic run, user
 \\
&&&
log,
commerci product, 
applic open, 
make applic, 
applic open sourc, \\
&&&
sourc product
\\

\hline
Specific& 5& GPL & 
gpl licens, 
gpl code, 
bsd licens, 
gpl softwar, 
gpl program, \\
licenses&&\& terms&
releas gpl, 
violat gpl, 
gpl incompat, 
program gpl, 
softwar gpl, \\
&&&project gpl, 
gpl licens code, 
gpl v2 licens, 
distribut gpl, 
term gpl
\\

\hline
Specific & 6& Various &
general public licens, 
gnu general public, 
public licens version,
\\
licenses && licenses &
patent licens,
lesser general public,
licens gnu general, 
gnu lesser 
\\
&&&general, 
applic web, 
patent claim, 
no-charg royaltyfre irrevoc, 
\\
&&&non-exclus no-charg royalty-fre, 
worldwid non-exclus no-charg, 
\\
&&&mozilla public licens,
affero general, 
licens grant
\\

\hline
Modifications& 7& Derivative&
deriv work, 
web servic, 
creat deriv work, 
work distribut, licens\\
&& \& warranties &
file complianc, 
version licens file, 
obtain copi licens, 
paid softwar, \\
&&&
cover work, 
origin work, 
work base, 
file complianc licens, \\
&&&basi warranti condit, 
condit kind express, 
warranti condit kind
\\

\hline
Specific & 8& Creative &
creativ common, 
cc by-sa, 
public domain, 
creativ common licens, \\
licenses && commons &
cc licens, 
licens creativ common, 
stack exchang, 
creativ common  \\
&&licenses&
attribut,
cc by-sa licens, 
licens materi, 
softwar author, 
licens \\
&&&
agreement, 
adapt materi, 
share alik, 
stack overflow
\\

\hline
Linking& 9& Static, &
open sourc librari, 
proprietari softwar, 
code open sourc, 
address \\
&&dynamic&
space, 
classpath except, 
program languag, 
dynam link, 
static 
\\
&& linking &
dynam, 
php code,
lt script, 
librari publish, 
static dynam link, \\
&&&
gpl classpath except,
project commerci, 
separ program
\\

\hline
Repositories& 10& Licenses \&&
pull request, 
releas code, 
licens code, 
code appli, 
piec code, 
code \\
&&  repositories&
public, 
github repositori, 
code snippet, 
publish code,
code project,\\
&&& 
code base, 
code github, 
fork project, 
origin author,
origin code
\\

\hline
Specific& 11& GPL/LGPL &
gpl librari, 
dynam link, 
lgpl librari, 
lgpl licens, 
librari licens,\\
licenses,&& \& libraries&
static link, 
link librari, 
librari gpl, 
share librari, 
code librari, \\
Libraries&&&3rd parti, 
librari releas, 
licens lgpl, 
parti librari, 
librari dll, 
\\

\hline
General OSS& 12& OSS &
open sourc project, 
open sourc licens, 
sourc licens, 
open sourc \\
&& products &
softwar, 
sourc softwar, 
licens open sourc, 
project open sourc, 
open 
 \\
&&&
sourc code, 
releas open,
make open, 
releas open sourc, 
free open  \\
&&&
sourc, 
softwar open sourc, 
allow chang,
make open sourc
\\

\hline
Specific& 13& Apache &
apach licens, 
licens file, 
sourc file, 
grant patent, 
apach softwar, \\
licenses&&license&
includ licens, 
apach licens version, 
file licens, 
notic file,
copi licens, \\
&&& 
file project, 
give recipi work, 
recipi work deriv, 
state chang, 
licens \\
&&&header
\\

\hline
General OSS& 14& FSF &
gnu general public, 
general public licens, 
free softwar, 
free  
\\
&&licenses&
softwar foundat, 
public licens publish, 
free softwar redistribut, 
\\
&&&
softwar foundat version,
foundat version licens, 
distribut hope, 
 \\
&&&
softwar redistribut modifi,
copi gnu general, 
licens publish free, 
 \\
&&&
modifi term gnu,
publish free softwar, 
redistribut modifi term
\\

\hline
Modifications& 15& Modifications&
provid sourc code, 
distribut sourc code, 
releas sourc code, 
make  \\
&& \& Distribution &
sourc code,
modifi sourc code, 
releas modifi, 
sourc code gpl,
publish\\
&&& sourc code, 
modifi version, 
program modifi version, 
releas modifi\\
&&&  version, 
sourc code modifi, 
sourc code icens, 
give sourc code,
gpl requir
\\

\hline
Commercial& 16& Selling &
softwar licens, 
licens softwar, 
piec softwar, 
develop softwar, \\
vs. OSS&& software &
offer sell, 
distribut softwar, 
softwar free, 
softwar develop, \\
&&&revers engin, 
softwar commerci, 
softwar compani, 
kind licens, \\
&&&licens restrict, 
softwar product, 
make softwar
\\
\hline
\end{tabular}
\end{adjustbox}
\end{table}

We grouped the 16 topics into the broader categories by considering the terms in each topic. We inspected manually 20 posts from each topic to ensure that the broader category is appropriate for the specific topic and present below these broader categories.

\textbf{Specific licenses.} This category gathers the largest number of topics (7 topics) and contains mainly questions about using software under a specific license, as well as questions on the license text, compatibility and version upgrading. \emph{Topic 1} refers to the MIT license and notice texts, as MIT appears in many terms and many terms refer to notices, i.e. copyright notices, permission notices and including notices. For instance, OS \#7203 is about using MIT licensed software without a copyright notice:\\
``\emph{[...]a few repositories do not have any copyright notice, while it is published under MIT license.[...] I wonder how I can follow MIT License. Should I write new copyright notice for [...] by myself? Can I skip including copyright notice of [...]? Or can’t I use this much popular library legally}."

\emph{Topic 3} refers primarily to GPL licenses, as its name and versions appear many times in the terms, whereas this may be discussed for specific products and their editions with possibly different licenses, including commercial ones, e.g. community edition. For instance, OS \#1021 refers to versioning:\\  
``\emph{How do I upgrade from GPLv2 to GPLv3?}"\\
, whereas SO \#3027364 refers to using old GPL-licensed code:\\
``\emph{Can I create a new project based on an abandoned one licenced under GNU GPL v2?}"

\emph{Topic 5} also makes many references to GPL. We find also words that refer to the GPL terms, violations of GPL and incompatibility to GPL. Therefore, the topic is more relevant to GPL and its terms. Law \#4592 refers to compatibility of GPL with BSD:\\
``\emph{[...]isn't gpl already incompatible with the 4-clause bsd licence that precedes the above comment, due to the advertisement clause?}"

\emph{Topic 11} concerns GPL and LGPL licenses and there is a reference to software libraries, as the text of LGPL indicates being used in software libraries\footnote{From the text of LGPL-2.1: A "library" means a collection of software functions and/or data prepared so as to be conveniently linked with application programs (which use some of those functions and data) to form executables.}. For instance, OS \#9599 refers to understanding LGPL text:\\
``\emph{What is the purpose of the LGPL re-linking requirement?}"

In \emph{Topic 6}, we find references to various licenses: GPL, LGPL, AGPL, MPL, that leads to the conclusion that posts in this topic refer to characteristics of specific licenses. For instance, about using a specific license in SEng \#179676:\\
``\emph{Can AfferoGPLv3 code be used in GPLv3 code?}"

\emph{Topic 8} contains terms that refer to creative commons licenses. Specific licenses are indicated, such as cc\_by-sa, whereas cc\_by-nc-sa also appears in the terms (after the first 15 most common terms). The presence of Stack Exchange and Stack Overflow in the list leads to the conclusion that some questions refer to how content from Q\&A sites should be used. Posts related to the public-domain can also be found. \emph{Non-source Code Licensing} is one of the licensing bugs categories mentioned in~\cite{vendome2018distribute}, although in this topics category we find mainly questions on two other licensing bug categories: \emph{Licensing Content} and \emph{Other Intellectual Property issues}. The following relevant example can be found in Law \#38324:\\
``\emph{Has the scope of Share Alike narrowed between CC 3.0 and 4.0?}"

Apache license is the main topic in \emph{Topic 13}. For instance, OS \#8819 refers to terms of the license:\\
``\emph{Does renaming compiled Java packages violate the Apache license?}"

\textbf{Conditions.} There is a very close meaning with the previous category, as different licenses contain various conditions. \emph{Topic 2} mentions copyright, but mainly contains terms about conditions, warranties, and liability with reference to specific software that leads to the conclusion that such aspects are discussed in the framework of specific products. For instance, SO \#23458414:\\
``\emph{How do I include django license in my project?}"\\
includes the whole text of the license in its body. The text of a draft license is also included in the body of SEng \#246845 with title:\\
``\emph{Is any one of these the right way to write a license for your opensource project forked and modified from an existing one with different authors?}".

\textbf{Commercial vs. OSS.} This category includes 2 topics. \emph{Topic 4} corresponds to commercial software and its source code, as we encounter the names of specific products. Posts in this topic may refer to the use of commercial software and opening the source code. For instance, SO \#1631177 asks whether the commercial license of MySQL needs to be purchased:\\
``\emph{MySQL: Do I have to purchase a license for InnoDB if I plan to use it in a for-profit setting (commercial use)?}"

\emph{Topic 16} has references to selling software and commercial software that points to posts related with how software products will be offered, as in OS \#8354:\\
``\emph{[...]I am creating a software project that relies on paid license software. I will release this software for free to help the community. But my users have to buy the software used in my project in order to get the benefit of my project. So kindly share your ideas about what kind of license I can add in this situation to my project?}"

\textbf{Modifications.} \emph{Topic 7} has many references to derivative works. Some terms relate also to warranties
. OS \#9535 discusses a specific case of a derivative work:\\
``\emph{[...]Does this imply that a recipient of GPLv3 software is legally allowed to circumvent technological measures placed by the redistributor to the extent that it interferes with the user's rights granted by the license?[...]}."

\emph{Topic 15} has many terms that refer to altering and distributing OSS. The Law \#39329 refers to making modifications public:\\
``\emph{Are there any AGPL-style licences that require source code modifications to be public?}"

\textbf{Linking.} Static and dynamic linking is discussed in \emph{Topic 9}. The kind of linking may define how a software is being used. OS \#4230 that contains a reference to GPL text states this issue:\\
``\emph{[..]Where's the line between two separate programs, and one program with two parts?[...]If the modules are included in the same executable file, they are definitely combined in one program. If modules are designed to run linked together in a shared address space, that almost surely means combining them into one program.[...]}."

\textbf{Repositories.} \emph{Topic 10} refers to using source code and online repositories, as there is a reference to GitHub, forks and pull requests. The issue of reusing source code from online repositories, or of the obligations when forking a repository concern practitioners. OS \#6623 asks how users will treat software uploaded in a GitHub repository:\\
``\emph{[...]I would like to put on GitHub the Perl/MySQL code that generates everything that the users see to allow those that want to make improvements on their own to do so and I can simply push them into production.[...]}", whereas SEng \#197410 refers to how the user that has modified a forked project should do:\\
``\emph{Accepted best practices for setup.py of a forked project}."

\textbf{General OSS.} In \emph{Topic 12}, ``open source" and relevant terms are repeated. It refers to general posts about open source software licenses, e.g. OS \#895:\\
``\emph{Does a license needs to be approved by the OSI to be an open source license?}"

\emph{Topic 14} contains many references to FSF (Free Software Foundation) and GNU that leads to the conclusion that it refers to licenses published by FSF. For instance, SO \#1486894 has a reference to GPL:\\
``\emph{Adding GNU GPL Licence to C\# App}," and SEng \#281949:\\
``\emph{How to properly apply a license}" uses GPL-3.0 in its body as example.

\subsubsection{RQ3. Which questions on OSS licensing get more attention and which seem} more difficult to answer?

In order to understand better the questions practitioners pose, we investigate in this section the attention received by questions that mention specific licenses or belong to licensing topics, we measure how long it takes for questions with specific licenses or on specific topics to get an accepted answer, and whether they get an accepted answer. As popularity indicators, we have relied on the following: number of posts, average views, average favourites and average score, that have been employed for this purpose also in previous works~\cite{bajaj2014mining,abdellatif2020challenges}. The views provide a measure for how many users or in how many cases the question was relevant to the user. The favourites provide a way to see in how many cases the practitioners thought the post would be useful for others. Score in Stack Exchange sites is calculated by using the number of up votes and down votes a post receives. 

Table~\ref{tab:licenses-popularity} shows the attention received by questions that contain references to the top 25 appearing licenses. Questions that include reference to the Apache license without version indication have the highest number of views followed by LGPL and BSD. The posts that mention the Apache license are limited and the highest number of views may be attributed to the fact that users may not have access to many other resources concerning the license (although there are many posts with reference to Apache-2.0). This may also be the case for the high number of average favourites for the MS-PL license. Unlicense, a license equivalent to the public-domain, has the same (also highest) number of average favourites and the highest average score for the posts that mention the license; therefore, the relevant posts may be considered a good source of information for this license that other users can also utilise. In the average score, the second highest value appears for questions mentioning WTFPL that may be dedicated to the uncommon name of the license. We observed that questions that mention WTFPL refer to using the license and clarifications for this use (e.g. Law \#104: ``\emph{Does the WTFPL legally disclaim warranties?}"). 

\begin{table}
  \caption{Licenses frequencies and popularity indicators.}
  \label{tab:licenses-popularity}
  \centering
  \begin{adjustbox}{width=0.8\textwidth}
  \begin{tabular}{crrrr}
    \hline
     \textbf{License} & \textbf{\#Posts} &\textbf{Avg. views} &\textbf{Avg. favourites}  & \textbf{Avg. score}\\
     \hline
GPL	&1533 & 2,112.5  &  1.4  & 5.7\\
MIT	&1047 & 1,722.5  &  1.3  & 5.6\\
GPL-3.0	&593 &1,399.8  &  0.9 & 5.0\\
LGPL	&525 & \textbf{2,600.7}  &  1.8 & 5.8\\
GPL-2.0	&450 & 1,562.5& 1.0 & 5.3\\
Apache-2.0	&413 & 1,488.2  &  0.9 & 4.8\\
AGPL-3.0	&241 & 1,634.1 &    1.0 & 4.7\\
public-domain	&175 & 1,092.0 &    1.0 & 5.6\\
BSD	&162 & 2,177.0 &    1.5 & 5.9\\
Apache	&108 & \textbf{3,688.8}  &  2.1 & 7.0\\
BSD-3-Clause	&87 & 1,348.7    &0.9 & 4.2\\
LGPL-2.0	&82 & 1,177.8 &    0.7 & 3.4\\
LGPL-3.0	&80 & 1,378.0 &    0.6 & 4.0\\
MPL	&65 & 1,336.9& 1.0& 4.8\\
LGPL-2.1	&64 & 1,206.2  &  0.7 & 3.5\\
EPL	&44 &  1,223.9& 1.0 & 4.8\\
BSD-2-Clause &40 & 2,116.9 &    1.6 & 5.9\\
MPL-2.0	&38 & 411.0 &  0.4 & 3.6\\
WTFPL	&38 & 1,097.2  &  1.3 & \textbf{8.8}\\
MS-PL	&36 & 2,090.7   & \textbf{1.9} & 6.7\\
zlib/libpng	&35 & 1,099.8  &  0.7 & 3.9\\
CDDL-1.0	&31 & 1,515.1  &  1.8 & 5.3\\
Unlicense	&26 & 1,744.6  &  \textbf{1.9} & \textbf{9.2}\\
GPL-3.0+	&19 & 440.4 &  0.2 & 3.5\\
GPL-2.0+	&17 & 404.8 &  0.6 & 3.6\\
X11	&17 &  991.0  &  1.1 & 4.8\\
  \hline
\end{tabular}
\end{adjustbox}
\end{table}

The frequencies of posts per topic and respective category and the average views, favourites and scores are shown in Table~\ref{tab:topics-popularity}. \emph{Specific licenses} is the most popular topic gathering 42.8\% of the total number of questions, followed by \emph{Commercial vs. OSS} in 16.4\% questions. The licenses that appear primarily in \emph{Specific licenses} topics are the top licenses present in the questions based on the license analysis (Table~\ref{tab:licenses-popularity}). Posts in the \emph{Commercial vs. OSS} category receive the highest average number of views and this is plausible, as this is a subject that concerns many developers either about using commercial software of making their software commercial or OSS, and it has been investigated also in the past~\cite{bitzer2004commercial}. The average views in \emph{Creative commons licenses} topic are lower than in other topics and this may be attributed to the fact that practitioners are not employing non-software licenses as much as software licenses. Considering the average score, users find the posts available for the MIT license \& notices, and OSS products useful, although posts that indicate only MIT have a lower average score (Table~\ref{tab:licenses-popularity}). 

\begin{table}
  \caption{Topics frequencies and popularity indicators.}
  \label{tab:topics-popularity}
  \centering
  \begin{adjustbox}{width=0.99\textwidth}
  \begin{tabular}{ccrrrrr}
    \hline
     \textbf{Category} & \textbf{Topic} & \textbf{\#Posts} & \textbf{Avg.} & \textbf{Avg.} & \textbf{Avg.}\\
     &&&\textbf{views}  &\textbf{favourites} & \textbf{score}\\
     \hline
Specific licenses&	MIT \& notices (1)&	338&	2,309.3&	1.6	&\textbf{6.6}\\
&	GPL (3)	&258&	1,325.3	&0.7&	4.2\\
&	GPL \& terms (5)&	465&	1,724.8&	1.1	&5.3\\
&	Various license (6)&	203&	2,384.9&	0.9&	4.9\\
&	Creative commons licenses (8)&	600&	953.3&	0.8&	5.0\\
&	GPL/LGPL \& libraries (11)	&494&	1,411.1&	0.9	&4.0\\
&	Apache license (13)	&505&	2,161.3&	1.2&	5.5\\
\hline
Commercial vs. OSS &	Commercial software (4)&	\textbf{695}&	\textbf{2,509.5}&	1.2&	4.1\\
&	Selling software (16)&	401&	\textbf{2,594.8}&	\textbf{2.1}&	6.2\\
\hline
Conditions&	License conditions (2)&	172&	1,575.5&	1.3&	5.8\\
\hline
Modifications&	Derivative \& warranties (7)	&187&	1,074.7&	1.0&	5.2\\
&	Modifications \& Distribution (15)&	615	&1,245.2&	0.8	&4.3\\
\hline
Linking&	Static, dynamic linking (9)&	190&	1,107.9&	0.6&	3.8\\
\hline
Repositories&	Licenses \& repositories (10)&	583&	1,281.5&	1.0&	5.1\\
\hline
General OSS&OSS products (12)&	\textbf{783}&	1,582.2&	\textbf{1.7}&	\textbf{6.5}\\
&	FSF licenses (14)&	94	&983.1&	0.6	&3.4\\
     \hline
\end{tabular}
\end{adjustbox}
\end{table}

We are also investigating indicators of difficulty of questions by examining whether the question has received an accepted answer and the median time to get an accepted answer~\cite{abdellatif2020challenges}. A question without an accepted answer or a question that takes a long time to answer may not be necessarily difficult, as this might be attributed to other factors, such as experts' availability in a given time period. We argue though that these metrics assist in examining how posts are being answered providing indicators for difficulty. We have performed this analysis based on the licenses that appear in the posts and on the questions topics.

In order to examine whether users receive an answer to their questions, we calculated the mean number of answers per license for the licenses that appear most often in questions. Overall, the mean number of answers a question receives regardless of the site and the licenses it contains is 1.7. The number of answers per license is relatively low (Table~\ref{tab:licenses-difficulty}). Although some license questions receive a large number of answers, e.g. questions with reference to GPL and LGPL receive up to 15, with reference to Apache-2.0 or MIT up to 10 and with reference to the public-domain up to 9, most questions receive only 1 answer, indicating that the users get an answer but usually they cannot choose from many answers. The average is higher for licenses that appear more often in questions, although the highest mean appears for GPL-2.0+ followed by WTFPL that appear in positions \#25 and \#19 in the questions respectively. 

Table~\ref{tab:licenses-difficulty} depicts also the median time to get an accepted answer for the question per license included in the question body. Questions with reference to WTFPL are answered faster than  others and this might be attributed to the short text of the license (5-6 lines) that does not state any real obligations or conditions for using the license. On the other end, questions indicating EPL and Unlicense take longer to answer appropriately. This may be attributed to the less frequent use of these licenses (e.g. EPL is not compatible to GPL), but it is unexpected for the case of Unlicense, since it points to the public-domain. The longest time to an accepted answer is found in the case of LGPL-3.0, although lower values are encountered for other versions and especially for license indications without version, which are more frequent. A similar observation can be made for GPL, where license indications without version and GPL-2.0/GPL-2.0+ receive an accepted answer faster in relation to GPL-3.0/GPL-3.0+. Questions that mention BSD-3-Clause and LGPL-3.0 have many questions without an accepted answer. By manually inspecting those questions, we observed that they refer to specific comparisons of the licenses (e.g. SEng \#114588: ``\emph{Downsides of GNU LGPL v3 vs. GNU LGPL v2.1?})," for applying it or to cases of software using the license (e.g. OS \#10287: ``\emph{Can an NPM package have an MIT license on github if it depends on MIT packages with BSD-2, BSD-3 and Apache 2 licenses})." There is a low ability of the community to answer such questions for BSD-3-Clause and LGPL-3.0, although the respective percentages of questions without accepted answer for close licenses (BSD-2-Clause, LGPL-2.0, LGPL-2.1) are lower. 

\begin{table}
  \caption{Indicators of difficulty of questions based on licenses that appear in the post.}
  \label{tab:licenses-difficulty}
  \centering
  \begin{adjustbox}{width=0.9\textwidth}
  \begin{tabular}{crrrr}
    \hline
     \textbf{License} & \textbf{\#Questions} & \textbf{\%Questions} & \textbf{Avg. \#} & \textbf{Median time to} \\
     &\textbf{without accepted}& \textbf{without accepted} & \textbf{answers} &\textbf{accepted (mins)}\\
     \hline
GPL & 534 & 34.8\%& 1.9& 100.5\\
MIT & 384& 36.7\%& 1.5& 195.7\\
GPL-3.0 & 227& 38.3\%& 1.6& 182.9\\
LGPL & 192& 36.6\%& 1.9& 115.8\\
GPL-2.0 & 160& 35.6\%& 1.7& 121.1\\
Apache-2.0 & 175& 42.4\%& 1.3& 223.3\\
AGPL-3.0 & 110& 45.6\%& 1.4& 261.0\\
public-domain & 62& 35.4\%& 1.9 & 192.7\\
BSD & 64& 39.5\% & 1.8& 106.2\\
Apache & 40& 37.0\% & 1.5& 182.7\\
BSD-3-Clause & 44& \textbf{50.6}\% & 1.1& 211.4\\
LGPL-2.0 & 28& 34.1\% & 1.4& 229.1\\
LGPL-3.0 & 38& \textbf{47.5\%} & 1.4& \textbf{319.5}\\
MPL & 28& 43.1\% & 1.9& 164.8\\
LGPL-2.1 &23 & 35.9\% & 1.3& 205.6\\
EPL & 17& 38.6\% & 1.5& \textbf{310.9}\\
BSD-2-Clause & 15& 37.5\% & 1.3& 280.3\\
MPL-2.0 & 15& 39.5\% & 1.1& 271.7\\
WTFPL & 11& 28.9\% & \textbf{2.0}& 114.8\\
MS-PL & 9& 25.0\% & 1.5& 114.8\\
zlib/libpng & 10& 28.6\% & 1.3& 218.2\\
CDDL-1.0 & 8& 25.8\% & 1.5& 187.0\\
Unlicense & 9& 34.6\% & 1.7& 288.3\\
GPL-3.0+ & 5& 26.3\% & 1.5& 260.17\\
GPL-2.0+ & 4& 23.5\%& \textbf{2.1}& 112.4\\
X11 & 5& 29.4\% & 1.9& 169.3\\
  \hline
\end{tabular}
\end{adjustbox}
\end{table}

The relevant indicators of difficulty concerning licensing topics are visible in Table~\ref{tab:topics-difficulty}. The questions on linking are the most difficult to answer (they have the highest percentage of posts without an accepted answer). The type of linking is not always easy to determine and relevant techniques are discussed also nowadays, e.g. for dynamic linking~\cite{gerbarg2020techniques}. Questions on derivatives and warranties take longer to answer than all other topics and this may be related with the difficulty in interpreting derivative works from a legal perspective\footnote{\url{https://www.linux.com/news/software-derivative-work-jurisdiction-dependent-determination/}}. Posts concerning repositories take less time to get an accepted answer, probably due to the wide spread of online repositories and the vast number of users that may be able to provide an answer. 

\begin{table}
  \caption{Indicators of difficulty of question licensing topics.}
  \label{tab:topics-difficulty}
  \centering
  \begin{adjustbox}{width=0.99\textwidth}
  \begin{tabular}{ccrrrr}
    \hline
     \textbf{Category} & \textbf{Topic} & \textbf{\#Questions} & \textbf{\%Questions} & \textbf{Avg. \#} & \textbf{Median time} \\
     &&\textbf{without}&\textbf{without} &\textbf{answers}& \textbf{to accepted}\\
     && \textbf{accepted} &\textbf{accepted} && \textbf{(mins)}\\
     \hline
Specific licenses&MIT \& notices (1) &	119& 35.2\% & 1.5 &184.2\\
&GPL (3)	&95 &28.1\% & 1.7 &78.5\\
&GPL \& terms (5)	&152 &32.7\% & \textbf{2.0} &110.5\\
&Various license (6)	&81 &40.0\% &1.6 &169.2\\
&Creative commons licenses (8)	&254 &42.3\% &1.5 &\textbf{193.2}\\
&GPL/LGPL \& libraries (11)	&170 &34.3\% &1.7 &145.5\\
&Apache license (13)	&185 &36.6\% & 1.4&161.2\\
\hline
Commercial vs. OSS&Commercial software (4)	&255 & 36.7\%&1.8 &91.9\\
&Selling software (16)	&175 &\textbf{43.6}\% &1.8 &139.2\\
\hline
Conditions &License conditions (2)	&70 & 40.7\%&1.6 &142.6\\
\hline
Modifications &Derivative \& warranties (7)	&71 & 38.8\%& 1.6&\textbf{254.1}\\
&Modifications \& Distribution (15)	&233 &37.9\% &1.7 &93.7\\
\hline
Linking & Static, dynamic linking (9)&84 & \textbf{44.2}\%& 1.6&154.2\\
\hline
Repositories &Licenses \& repositories (10)	&207 & 35.5\%& 1.8&106\\
\hline
General OSS&OSS products (12)	&265 & 33.8\%& \textbf{2.1}&103.8\\
&FSF licenses (14)	&33 &35.1\% &1.8& 85.5\\
  \hline
\end{tabular}
\end{adjustbox}
\end{table}

\subsubsection{\textbf{RQ4}. Which type of questions do OSS users ask about licensing?}

In order to examine the type of questions, we use a simple approach of searching the question word in the question title and body. In order to ensure that the \emph{which} question is identified correctly, we manually inspected all cases and performed appropriate adaptation when `\emph{which}' was used instead in a relative clause in the text (which is very usual also in the text of licenses). We have used the questions of all sites together. 

The type of questions based on the categories identified in the topic modelling analysis are shown in percentages in Table~\ref{tab:questions-type}. The largest number of questions cannot be classified to one of the common question types (\emph{Other} questions), which shows that users are using more complex ways to formulate their question. \emph{Why} questions appear less often, whereas \emph{What} questions are the most common.  

\begin{table}
  \caption{Licensing question types percentages on Stack Exchange sites.}
  \label{tab:questions-type}
  \centering
  \begin{adjustbox}{width=0.84\textwidth}
  \begin{tabular}{cccccc}
    \hline
     \textbf{Category} & \textbf{\% How} & \textbf{\% Why} & \textbf{\% What} & \textbf{\% Which} & \textbf{\% Other}\\
     \hline
    Specific licenses&	7.4 &	1.1&	13.4&	2.5& 75.6\\
Commercial vs. FOSS&	8.3 &	0.6&	12.9&	2.7& 75.5\\
Conditions	&6.4&	1.7&	13.4&	2.3&	76.7\\
Modifications&	6.1&	1.9&14.5&	3.5&74.1\\
Linking&	5.8	&1.6&	12.6&	2.1	&78.4\\
Repositories	&7.2	&0.5&	18.4&	4.3&	70.7\\
General OSS&	7.2&	1.3&	17.2&	1.7&72.6\\
\hline
\textbf{ALL}&	7.3	&1.1&	14.4&	2.6	&74.7\\
  \hline
\end{tabular}
\end{adjustbox}
\end{table}

A number of examples for each question type using top score questions are provided in Table~\ref{tab:type-examples}. In \emph{How}, users are asking how to perform an action concerning a license or a software. In \emph{Why}, users ask different questions about why a license is used and why specific choices about licensing are usually made. In \emph{What}, users are asking about specific licenses and their differences. Finally in \emph{Which}, questions on compatibility and license selection are found. The manual verification of the \emph{Which} questions showed that the majority refers to recommendations about which license to use. \emph{Other} questions cover a variety of areas. We manually inspected 10 random such questions from each site. In the Law site, 345 such questions can be found. Most refer to clarifications about licenses and legal aspects of software licenses (e.g. finding legal experts) by using mainly 'can', 'may' question types. The Open Source site has 1,630 other questions. We did not encounter any specific question types among the questions we examined. 812 \emph{Other} questions are encountered in Software Engineering site. Some questions present a hypothetical situation using 'if' or ask if they can do something using different terminology, e.g. 'can', 'is it possible', 'is it wrong', 'does it violate'. Finally, Stack Overflow has 796 other type questions. Some posts on whether the user can do something are also encountered here, e.g. 'is it legal'.

\begin{table}
  \caption{Licensing question types examples on Stack Exchange sites.}
  \label{tab:type-examples}
  \centering
  \begin{adjustbox}{width=0.9\textwidth}
  \begin{tabular}{l}
    \hline
     \textbf{How} \\
    \emph{How do I protect [...] code?} (SO \#261638) \\
    \emph{How to add license[...]} (SO \#31639059)\\
    \emph{How to change the license[...]} (SO \#20243214)\\
    \emph{How to manage a copyright notice[...]} (SEng \#157968)\\
    \emph{How can free and open source projects be monetized?} (OS \#88)\\
    \hline
    \textbf{Why}\\
    \emph{Why do people pick the MIT license over BSD[...]?} (SEng \#82321) \\
    \emph{Why is CC BY-SA discouraged for code?} (OS \#1717) \\
    \emph{Why does Linux still use the GPLv2?} (OS \#1774) \\
    \emph{Why is there no public domain licensing in Europe?} (OS \#9871) \\
    \emph{Why is the warranty disclaimer section of a licence [...] shouted?} (SEng \#36513) \\
    \hline
    \textbf{What} \\
    \emph{AGPL - what you can do and what you can't} (SEng \#107883) \\
    \emph{What are the differences between GPL v2 and GPL v3 licenses?} (SO \#41460) \\
    \emph{What are the real life implications for an Apache 2 license?} (SEng \#56927) \\
    \emph{What can I assume if a publicly published project has no license?} (OS \#1720) \\
    \emph{What license must I use on[...]?} (OS \#27) \\
    \hline
    \textbf{Which} \\
    \emph{Which open source licenses are compatible with[...]} (SO \#459833) \\
    \emph{Is there a chart of which OSS License is compatible with which?} (SO \#1978511) \\
    \emph{Which is the most permissive open-source license?} (SEng \#134687) \\
    \emph{Which Google Code license should I use?} (SO \#1082101) \\
    \emph{Which licenses are free for commercial use?} (SEng \#134802) \\
    \hline
    \textbf{Other} \\
    \emph{Can public domain content be licensed?} (Law \#11015) \\
    \emph{MVN repository artifact licence?} (OS \#8801) \\
    \emph{Is source code required to be published if a program embeds GPL software?} (OS \#6681) \\
    \emph{If I use .Net Core 2.0 my software must by open software?} (SEng \#379366) \\
    \emph{GNU GENERAL PUBLIC LICENSE software as part of commercial product} (SO \#901131) \\
  \hline
\end{tabular}
\end{adjustbox}
\end{table}

\section{Discussion and Implications}

\subsection{Discussion}

\textbf{License indications in questions.} Our results show that the top license users mention in their posts is the popular strong copyleft license, the GPL. This may be either attributed to the fact that the GPL license is used very often in OSS projects, as GPL-3.0 and GPL-2.0 are used in total in 25\% of projects on the Black Duck Knowledge Base\footnote{\url{https://web.archive.org/web/20170824071559/https://www.blackducksoftware.com/top-open-source-licenses}} in 2017 or that the text of the license, and especially of GPL-3.0, is long and users need clarifications. Users may thus, be wondering how to treat software that carries GPL so that no violations can occur, how to correctly use the license in their works, or may compare the license with other licenses. The second top appearing license, the MIT, has a very short text and there is a tendency to find it very often, especially in the last years: it is the top license in the Black Duck Knowledge Base in 2017 (found in 32\% of projects). Apache-2.0 that has a relatively long text is also commonly found in the questions of our dataset (top 6th position) and is 3rd top license in the Black Duck Knowledge Base in 2017. Popular licenses, such as GPL, LGPL, BSD, are mentioned more often without version indication. The reason for this may be that practitioners do not realise that these are families of licenses and treat them as a single license. Practitioners may also have a version in mind, e.g. GPL-2.0 when they mention GPL or BSD-3-Clause when they mention BSD, and future work might be needed in order to examine it more thoroughly.

Most of the top licenses in our results (Table~\ref{tab:licenses-qa-sites-freqs}) have been approved by OSI, excluding WTFPL, CPOL-1.02 and OpenSSL license. The public-domain and creative commons licenses are also not present in OSI, as they have a different meaning from OSS licenses, but we have included them in our results due to the existence of software applications under the public-domain and of non-software resources licensed under creative commons. 

\noindent\fbox{%
    \parbox{\columnwidth}{%
There is an increase in the number of times permissive licenses are referred to in questions in 2015 and afterwards (in 2015-2020, Apache and its different versions, BSD and its different variants, MIT, and zlib/libpng are mentioned 23.1\% of all license indications compared to 8.0\% in 2008-2014) showing a higher interest toward permissive licenses that has also been observed in other studies\footnote{\url{https://resources.whitesourcesoftware.com/blog-whitesource/top-open-source-licenses-trends-and-predictions}}~\cite{vendome2017license}. There is also a tendency to ask about a larger number of licenses since 2015, indicating that less popular licenses are also considered for use or are examined by practitioners or are used in software the practitioners are using.
    }%
}

\textbf{Licensing topics in questions.} Comparing our results with the results of the topic modeling in the initial version of our work, we see that common categories and topics can be found (e.g. specific licenses, general OSS, linking, derivative and modifications), whereas repositories is a new topic that may be linked with the addition of SO to the dataset~\cite{kapitsaki2020developers}. 

\begin{figure}[!t]
\centering
\includegraphics[scale=0.42]{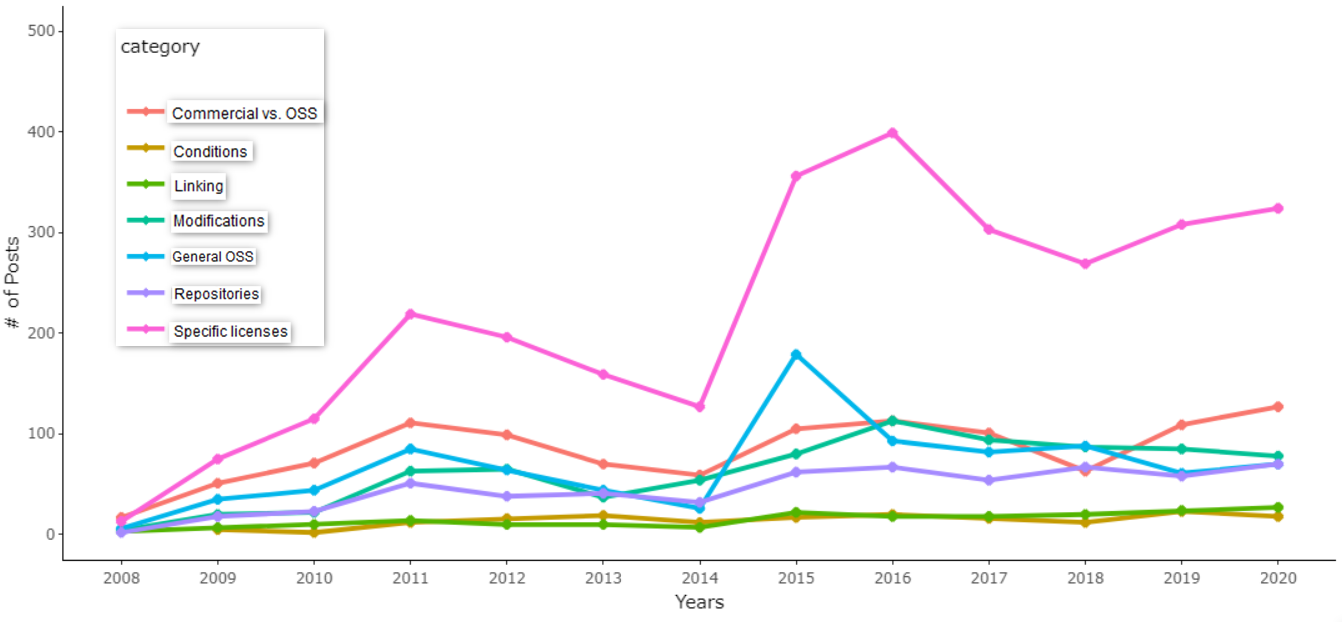}
\caption{Licensing topic categories evolution over time.}
\label{fig:mean-answers-per-license}
\end{figure}

Not all licensing topics appear with the same frequency in users' questions. \figurename~\ref{fig:mean-answers-per-license} depicts how the popularity of the broad categories changes over time, showing the number of posts that have a reference to each category. \emph{Specific licenses} is the most popular category and there is a significant increase in the number of posts in 2015 that can be seen till 2020 with a peak in 2016. The increase in 2015 is attributed to the creation of the Open Source site that year that gave users the opportunity to pose their questions. A similar increase is visible also in other topics, e.g. \emph{General OSS}. \emph{Conditions} and \emph{Linking} categories have less popularity and less changes over the years. \emph{Linking} is a more specific issue in open source licensing and this might explain the low numbers for this category that is also more difficult to answer, whereas \emph{Conditions} category has a large overlap with the \emph{Specific licenses} category. 

There is a sharp increase in the \emph{General OSS} category in 2015 that may be attributed to the creation of the Open Source site, but the sharp subsequent decrease in 2016 leads to thinking that the presence of the site as a new source to seek information for OSS may help practitioners get an answer to their questions without the need to pose a new question themselves. In the beginning of 2017 the GitHub open source guide was published\footnote{\url{https://github.com/github/opensource.guide}} and the decrease in posts in almost all categories in this year may be attributed to this to some extent, as it provides a new source of resources for OSS practitioners. 

\begin{figure}[!t]
\centering
\includegraphics[scale=0.36]{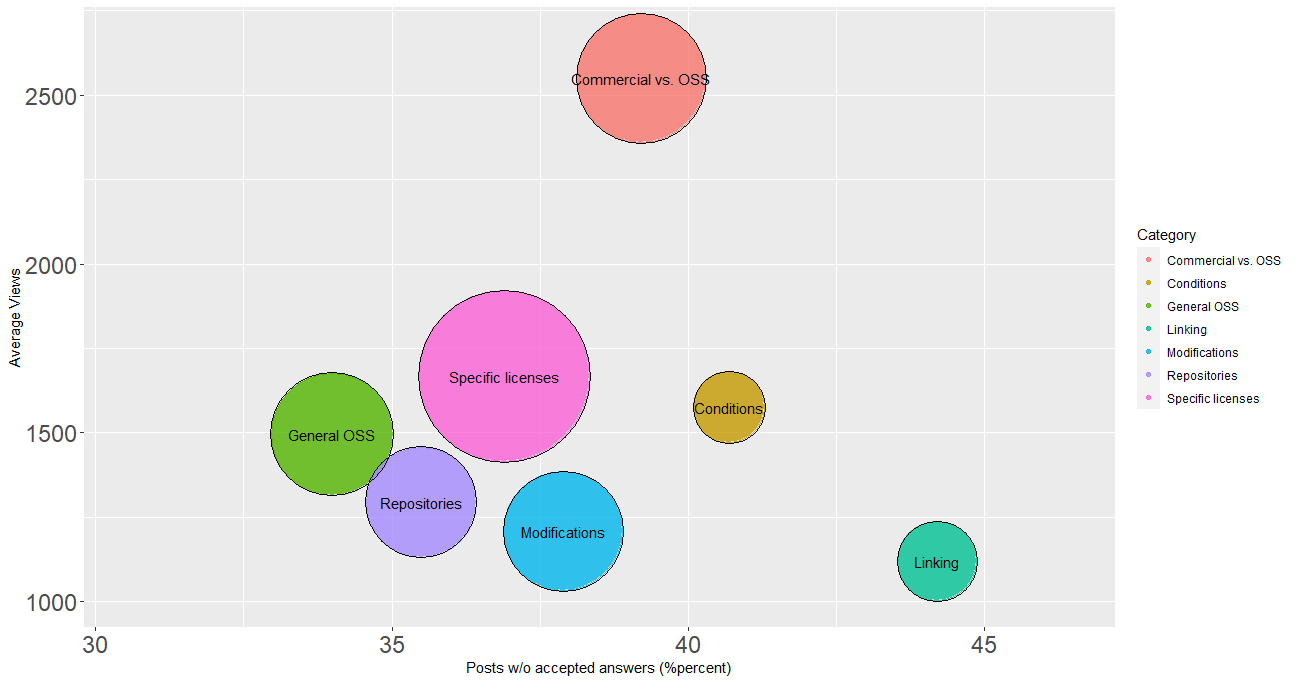}
\caption{Licensing topic categories difficulty versus popularity.}
\label{fig:bubble-plot}
\end{figure}

The attention received differs among topics, whereas a significant number of questions does not receive an accepted answer (37.7\% considering the whole question dataset). \figurename~\ref{fig:bubble-plot} shows the relation between the average views and the posts without an accepted answer for the topic categories. The bubble is proportional to the number of posts in the category. It verifies what we have already noted, e.g. that posts on \emph{Linking} get more attention but seem at the same time difficult to answer. Also \emph{General OSS} posts get less attention than some other categories but can be answered easier by Stack Exchange users, potentially due to their more generic nature. It is interesting to note that questions on selling software (in the \emph{Commercial vs. OSS} category) get a lot of attention but have at the same time a large number of unanswered questions, indicating that cases about deciding to sell software or using commercial software concern many users but are not easy to handle.     

\noindent\fbox{%
    \parbox{\columnwidth}{%
Most popular questions mention Apache and LGPL licenses and concern commercial software versus OSS. Questions on derivative and warranties, creative commons licenses and MIT, as well as questions that mention EPL without version and LGPL-3.0 are more complicated to answer or not many users of Stack Exchange sites know relevant issues well, as it takes longer to get an accepted answer. 
    }%
}

\subsection{Main lessons learned}

The results of our study can be useful for different kind of users. 

\textbf{Target group 1:} Educators and employers.

\textbf{Implications:} For \emph{educators and employers} the results of \textbf{RQ1} can point them to which licenses to focus on for adapting the training of software engineers, preparing them more efficiently for the software industry for using open source software. For MIT, practitioners ask more about using the license, e.g. ``\emph{Can I sell my MIT-licensed software if it is modified by others?}" (OS \#1230), ``\emph{If I license software under the MIT license, will I be able to collect patent royalties?}" (OS \#1398), about using code under the MIT license, e.g. ``\emph{Can I use MIT licence plugins in my commercial web site?}" (SEng \#103273), but this can be encountered for other licenses also. Therefore, educating engineers on main rights and obligations each license brings, for instance, based on each category as permissive or copyleft, is important. \textbf{RQ3} showed that some licenses, such as BSD licenses, LGPL licenses and Apache, receive less answers so authoring more educational content and chapters on these licenses may be helpful for practitioners. There is an overlap of the above comment with the results of the topic modeling of \textbf{RQ2}, as many topics (7 topics) refer to specific OSS licenses. But general aspects of licensing, such as licensing conditions (1 topic), creating derivative works (2 topics) and handling in online repositories (1 topic), can be considered important elements of an advanced software engineering course. The meaning of linking (1 topic) is also important. As a small number of relevant posts has been found in the Law site, collaboration with law experts would be useful, in order to gain a basic understanding of copyright. 

\textbf{Target group 2:} People who write licenses, such as legal experts who are thinking about writing new licenses, updating existing licenses or creating and improving existing documentation.

\textbf{Implications:} Our results indicate on which licensing
aspects legal experts should focus on, updating the documentation of existing licenses or software to address issues developers face. 
According to the results of all RQs and especially \textbf{RQ1}, \textbf{RQ2} and \textbf{RQ3}, developers are not certain about compatibility of a given license with other licenses, as a large number of questions concerns combinations of licenses, e.g. ``\emph{Can I license a program under Apache License 2.0, if it is using/including Eclipse Link (EPL or EDL)?}" (Seng \#180233) or ``\emph{How to use GPL v3 with Apache License 2.0?}" (Seng \#197710). Also many users are unsure about the general meaning of the license, e.g. ``\emph{LGPL/public-domain equivalent of Apache log4cxx?}" (SO \#5180367). The above indicate that license compatibility with uses of different licenses in the same software product should be part of a license documentation and needs to be specific on which the other licenses might be. The GNU maintains a page with information about compatibility of various licenses with the GPL licenses\footnote{\url{https://www.gnu.org/licenses/license-list.en.html#Introduction}}. This information is very useful for practitioners. Our results indicate that a number of licenses get more attention when they appear in posts in Stack Exchange sites, i.e. more views, such as LGPL and BSD, that may indicate that more explanations are needed for using these licenses or for software that carries these licenses. 

\textbf{Target group 3:} Software developers.  

\textbf{Implications:} Developers contributing to open source software can understand better the importance of licensing and the issues they need to be cautious about when reusing source code. Developers releasing their original work as open source can understand better the problems the future users of the software will have in respect to its licensing and the creation of derivative works using their software. Our results in \textbf{RQ1} show a high interest in MIT. MIT is used in many projects and, although it is a license with a short text, many developers are asking questions about how to use MIT-licensed software correctly or how to combine it with other licenses. Therefore, even when using a license with a short text, clarifications are needed and a large number of them are not easy to answer. Developers need to place information about where to get additional resources for a specific license inside the product that is being distributed, e.g. links to compatibility information, in order to make the life of licensees easier.

\section{Threats to validity}

In this section, we discuss threats that can affect the validity of our study and measures taken to mitigate them. 

\textbf{Internal validity.} Our research is built upon two assumptions: first, that the questions asked reflect the need for information of those who post them; and second, that most of the answers are relevant to the question posed (even if they do not provide a correct answer). Our manual verification of the data we collected for this study attest to the validity of both of these assumptions (as we describe below, we took further steps to make sure we remove from consideration non-relevant questions and answers). While Stack Exchange and similar Q\&A sites are widely used by software engineers, we have no evidence that they are used by lawyers. Thus we can only claim that are results reflect the information needs of software developers and not of law professionals.

We only used data regarding the question explicitly being asked, not how many people have the same question (and benefit from one that is already posted and answered---as opposed to asking that question again), therefore we do not have an indication of what questions are more important to developers or whether questions that were first asked several years back continue to be popular or now. We encourage future research to look into these issues.

\textbf{Construct validity.} It measures the degree to which we measure what we are supposed to measure. The dataset quality is a crucial aspect and may have affected our results. We have relied on existing practices for collecting our dataset (i.e. Stack Exchange API) and have relied on the use of large number of licensing relevant keywords and tags to identify questions and answers relevant to our study. We have also relied on extended automatic and manual filtering to remove noise from our collected dataset.

In order to limit the threat of including duplicates, we detected if the same body text exists in more than one questions in the different Stack Exchange sites. The first stage of preprocessing contained the removal of duplicates using the question id from every Stack Exchange site. However it is possible that the same user has posted the same question in more than one sites. Therefore, the second stage consisted in finding if the text of every question across the different Stack Exchange sites is exactly the same. Using this approach we removed 28 cases. In the scenario where different users post a question that was asked in the past, every site provides to the members the chance to report this (the report is a mark in the title of the question which is characterized as duplicate). Moderators detect and close duplicate questions and provide the link of the same previous posted question. The Q\&A site redirects users to the initial question and both API and search of Stack Exchange do not show the duplicates.

We do not expect that the filtering we have performed based on keywords may have affected our results removing some meaningful cases, as we manually verified the removal of many posts. In the case of Stack Overflow, we have used licensing-relevant tags but we expect that noise should have remained in our data, since licensing questions are outside the scope of Stack Overflow. We have manually filtered the SO posts to minimise this threat. We also filtered out a large number of questions from the Law Stack Exchange site, including posts with country or state names. Although this decreased the size of the Law dataset, it removed a large amount of noise.

A minor threat is that we have used surrogate metrics to measure questions' difficulty (i.e. median time to accepted answer), as this time may be affected by other parameters (e.g. experts' availability). Nevertheless, this is the only metric that can be considered for this measurement and has been applied in the literature in the past~\cite{abdellatif2020challenges}.

\textbf{External validity.} It refers to the extent we can generalise our findings~\cite{yin2013case}. With the goal of getting a wide perspective, we mined questions from four different Stack Exchange Q\&A sites, including those oriented towards questions regarding software engineering and law. In all these sites we found the results to be similar. Nonetheless, there exist several threats to validity that affect our generalisation. First, our sources are public and rely on volunteers for answers, thus it is possible that many of these questions are posed by developers working outside an organisation and without formal legal advice (i.e. hobbyists who develop open source software in their spare time). This has two implications: on the one hand, that questions posed might relate primarily to building non-commercial products, and on the other hand, the type of person asking the questions in a public forum is different than people working for an organisation that has formal legal advice. Second, using difference sources of information (such as mailing lists) might lead to different results. Thus, we cannot claim that our results are generalisable to any software engineering setting. Nonetheless we believe that developers who asked questions in these sites are likely to have a similar background to those who have legal advice, as a consequence, our results reflect common questions, concerns and misunderstandings that software developers in general have regarding software licensing.

\textbf{Conclusions validity.} It is the degree to which conclusions we reach about relationships in our data are reasonable. We examined different sources to collect our data. Although each site has slightly different audiences in mind, there is some consistency in the results, increasing the reliability of our conclusions that software developers phase the issues we discuss.

\section{Conclusions}

In this work, we have presented a study on the analysis on questions about open source software licensing in Stack Exchange sites (Stack Overflow, Software Engineering, Open Source and Law). We have examined the top licenses that appear in the posts and the licensing topics, whereas we have also observed how these evolve over time, the attention they receive and indicators of difficulty, and the type of questions asked. We observed that most of the questions are about specific OSS licenses, whereas the licenses mentioned in posts that get more attention are among the top licenses used in OSS products. Our results can be a useful source of information for educators and employers, license authors and developers in either better educating young developers or in being proactive in addressing practitioners needs on OSS licensing. 

As part of our future work, we intend to add more sources of information in order to draw more general conclusions that include also the usage of licenses, such as mailing lists of specific open source communities, existing documentation and question and answers on the websites of specific licenses, as performed in other works that combine data~\cite{zagalsky2018r}. We also aim to employ additional Natural Language Processing techniques, in order to better understand the texts relevant to licensing, e.g. via pattern matching. Repeating the study in 5 years, in order to examine what will have changed in that period, as new licenses emerge or other become more popular would be interesting as well.

\section*{Acknowledgment}
This research is co-financed by Greece and the European Union (European Social Fund- ESF) through the Operational Programme ``Human Resources Development, Education and Lifelong Learning" in the context of the project “Strengthening Human Resources Research Potential via Doctorate Research” (MIS-5000432), implemented by the State Scholarships Foundation (IKY)

\bibliography{sample-base}

\end{document}